# Measurements of a 2.1 MeV $H^-$ beam with an Allison scanner

C. Richard*
*Michigan State University, East Lansing, MI 48824, USA*

M. Alvarez, J.P. Carneiro, B. Hanna, L. Prost, A. Saini, V. Scarpine, A. Shemyakin
*Fermi National Accelerator Laboratory, Batavia, IL 60510, USA*

Transverse 2D phase space distribution of a 2.1 MeV, 5 mA $H^-$ beam is measured at the PIPIT test accelerator at Fermilab with an Allison scanner. The paper describes the design, calibration, and performance of the scanner as well as the main results of the beam measurements. Analyses of the recorded phase portraits are performed primarily in action-phase coordinates; the stability of the action under linear optics makes it easier to compare measurements taken with different beamline conditions, e.g. in various locations. The intensity of a single measured point ("pixel") is proportional to the phase density in the corresponding portion of the beam. When the Twiss parameters are calculated using only the high-phase density part of the beam, the pixel intensity in the beam core is found to be decreasing exponentially with action and to be phase-independent. Outside of the core, the intensities decrease with action at a significantly slower rate than in the core. This 'tail' comprises 10-30% of the beam, with 0.1% of the total measured intensity extending beyond the action 10-20 times larger than the rms emittance. The transition from the core to the tail is accompanied by the appearance of a strong phase dependence, which is characterized in action-phase coordinates by two 'branches' extending beyond the core. A set of selected measurements shows, in part, that there is no measurable emittance dilution along the beam line in the main portion of the beam; the beam parameters are practically constant over a 0.5 ms pulse; and scraping in various parts of the beam line is an effective way to decrease the transverse tails by removing the branches.

## I. INTRODUCTION

Fermilab is developing the Proton Improvement Plan, Stage Two (PIP-II), a program of upgrades for its injection complex [1]. The central part of the PIP-II project is an 800 MeV, 2 mA CW - compatible superconducting (SRF) $H^-$ linac, envisioned to be working initially in a pulsed mode for injection into the existing Booster synchrotron. A prototype of the PIP-II linac front end called PIP-II Injector Test (PIP2IT) [2] is being built to retire technical risks associated with acceleration at low energies and to demonstrate a capability to create an arbitrary bunch structure [3]. As of now, the Warm Front End (WFE) - a 15 mA DC, 30 keV $H^-$ ion source, a 2 m-long Low Energy Beam Transport (LEBT) and a 2.1 MeV CW RFQ, followed by a 10 m Medium Energy Beam Transport (MEBT) - has been assembled and commissioned [4] (Fig. 1). Eventually PIP2IT will also comprise two cryomodules accelerating $H^-$ ions up to 25 MeV.

As for most recent ion accelerators at low energy, the WFE houses multiple diagnostics to control and characterize the beam. In particular, means to measure the phase space distribution(s) of the beam are almost always present in order to better understand possible issues with losses from tail particles and matching in the subsequent accelerating structures.

In the PIP2IT MEBT, the transverse rms beam characteristics are reconstructed primarily through 1-D beam profiles recorded with scrapers and quadrupole scans [5], while information about details of the ion distribution in phase space are extracted from measurements with an Allison scanner. The latter measurements are the main subject of this paper, which is organized as follows.

Section II is devoted to the design of the Allison scanner, reasoning for choosing this device for measurements at the relatively high beam energy of 2.1 MeV, and features of its operation. Section III describes the MEBT beam line where the measurements were performed. Calibration of the scanner and performance tests are the subject of Section IV. Section V introduces the description of the particles' distribution in action – phase coordinates and the corresponding method of dividing the distribution into core and tails. A selection of measurements analyzed in action – phase coordinates are described in section VI, followed by a conclusion.

## II. ALLISON SCANNER

### A. Working principle

An Allison scanner, first proposed in 1983 [6], consists of two slits mounted on a common frame, two electric deflection plates between the slits, and a suppressor electrode behind the rear slit followed by a Faraday cup (collector in Fig. 2). The scanner measures the beam density distribution in phase space by stepping the assembly through the beam and scanning the voltage on the plates at each step. The position coordinate of a portrait pixel is defined by the position of the front slit; the voltage between the plates determines the pixel's angular coordinate; and the Faraday cup current at that given position and plate voltage is the pixel intensity. The output pixel matrix is referred in this paper as a phase portrait. Characteristics of the PIP2IT MEBT scanner are listed in Table I.

* richard@nscl.msu.edu

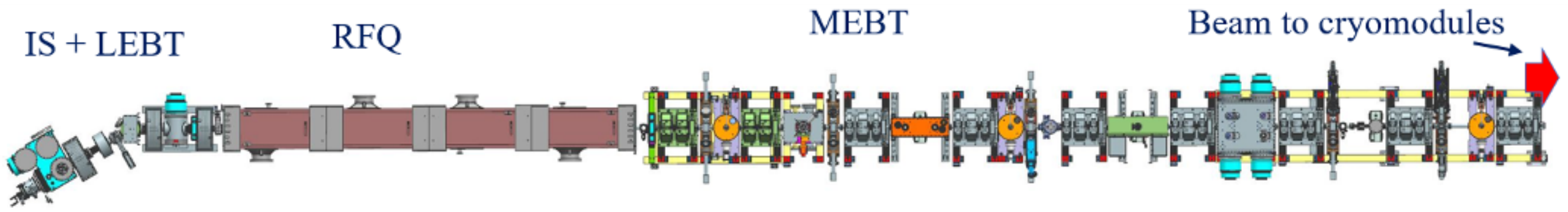

FIG. 1. PIP2IT warm front end (top view)

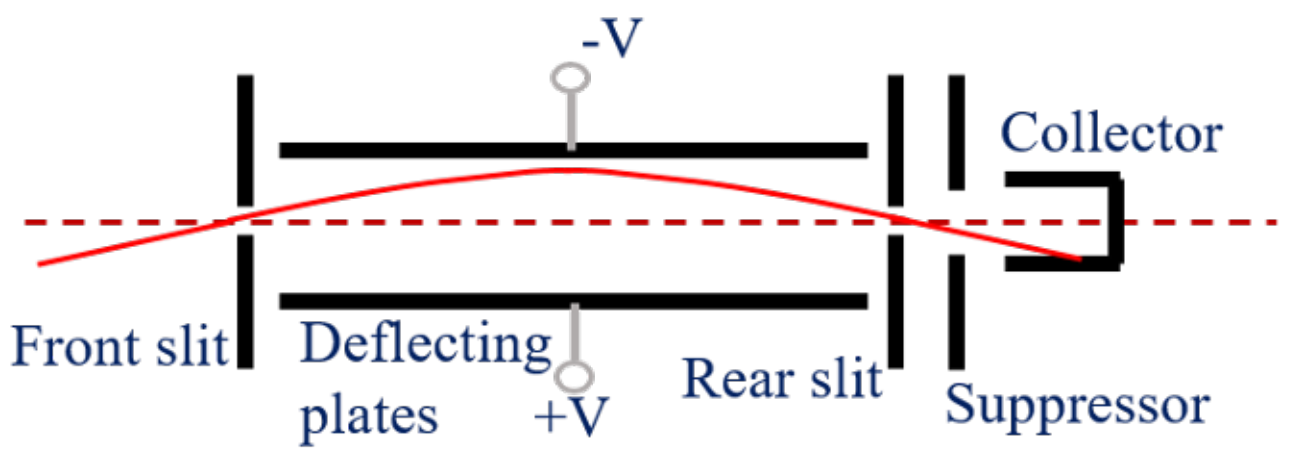

FIG. 2. Allison scanner schematic. The solid red line shows a beamlet trajectory.

TABLE I. Parameters of the PIP2IT MEBT scanner

| Parameter | Value | Unit |
|---|---|---|
| Slit size | 0.2 | mm |
| Slit separation | 320 | mm |
| Plate voltage | ±1000 | V |
| Plate length | 300 | mm |
| Plate separation | 5.6 | mm |
| Maximum measurable angle at 2.1 MeV | 12 | mrad |

Typically, Allison scanners are used for characterization of ion beams at energies of tens of keV (e.g. [7], [8]), which is also the case for the PIP2IT LEBT Allison scanner [9]. In the MeV-range, most ion accelerators to date use the so-called double-slit (e.g.: J-PARC's Temporal Beam Diagnostic system [10]), slit-grid (e.g.: CERN's Linac 4 diagnostics bench [11]) or slit-wire (e.g.: SARAF's Diagnostic plate [12]) configurations, where both the position and angle coordinates are determined by the positions of the slits, wires and/or grids. The authors are only aware of one notable exception: a "sweep plate emittance scanner for high-power CW ion beams" developed by the Advanced Technology and Development Center of Northrop Grumman, which was used on their "1.76 MeV Pulsed Beamline" in the mid-nineties [13, 14].

One concern that discourages using Allison scanners and other intercepting devices at higher energies is the beam-induced thermal stress that restricts studies of long-pulsed beams. This restriction is less important in the case of PIP2IT because of the special scheme [15] used in the LEBT, which provides nearly constant parameters through the entire pulse. Therefore, evaluation of the beam parameters with short pulses ($\sim 10\ \mu$s) represents well the entire range of interesting time scales.

Another challenge is the decrease of the deflection angle with ion energy $E_i$ for a given electric field (as $1/E_i$). However, countering this effect is a decrease of the typical angular range in the beam with acceleration. For the same Twiss parameters and normalized emittance, the angles drop as $1/\sqrt{E_i}$. In addition to decreasing the required angular range of the device, it allows for placing the deflection plates closer, thus increasing the electric field for a given voltage. The minimum effective gap between the plates is determined mainly by the height of the parabolic trajectory between the plates, which is proportional to the angle to be measured. By adjusting the gap, an Allison scanner can be adequate for higher energies until the slit size and achievable alignment errors become comparable with the gap size.

There were multiple factors that eventually determined the choice of the Allison scanner as the main tool for measuring the phase portraits at PIP2IT, including speed of acquisition and accuracy. However, the most decisive of them was the very positive experience with the LEBT Allison scanner [9] and possibility to reuse its electronics and software. Also note that all the emittance measuring devices at the facilities listed previously were implemented as temporary instruments (on a specially designed diagnostics station) and are not included in the operational configuration of those beam lines. The PIP2IT MEBT Allison scanner was designed to eventually permanently stay after first two quadrupole doublets.

### B. Mechanical design

The design of the MEBT scanner (Fig. 3) is based on the PIP2IT LEBT Allison scanner, which, in turn, is a modified version of the SNS Allison scanner [8]. Its front slit plates, made of a molybdenum alloy known as TZM, are bolted to a stainless steel water-cooled block with a graphite foil wedged between them. For a beam with a relatively low peak power density (<200 W/cm$^2$ in $\mu$s scale), the slits can withstand up to 100 W of average beam power.

To avoid angular restriction of the passed beamlet by the slits as reported in Ref. [16], the slit restricting surfaces are manufactured as knife edges with radius not exceeding 0.13 mm (see insert in Fig. 3). The estimated maximum reduction in pixel intensity due to the slit thickness [16] is approximately 1%, so this correction was neglected in all analyses.

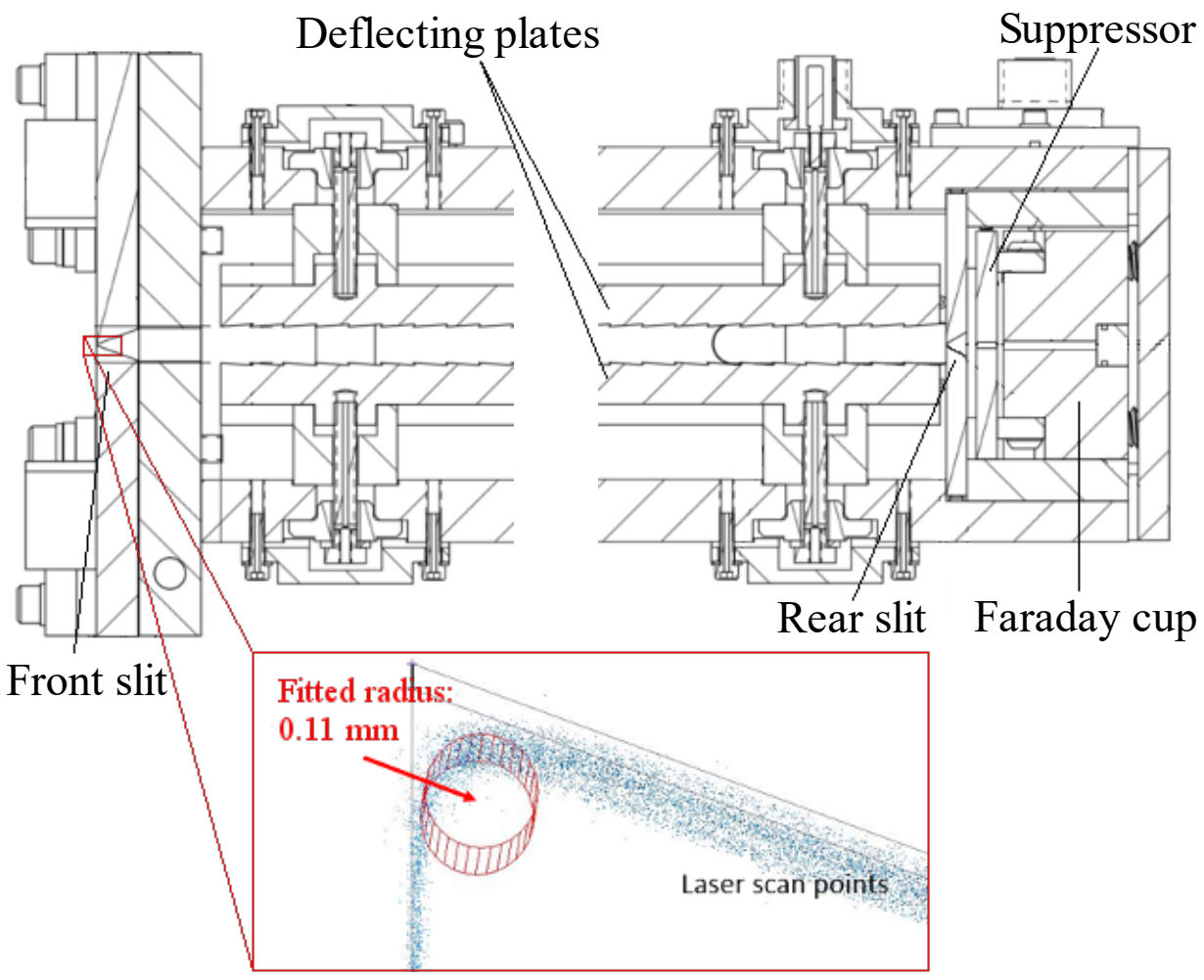

FIG. 3. Front and rear slits of the Allison scanner. The radius of the slits' knife edge was determined by laser scan.

The deflection plates are aligned parallel, so that the gap along the scanner stays constant within ±25 $\mu$m. The suppressor electrode and the Faraday cup are implemented as slits with openings of 0.89 mm and 1.27 mm, correspondingly. Typically, the suppressor electrode is biased at -100 V repelling secondary electrons both exiting from the Faraday cup and coming in the opposite direction from the rear slit.

The scanner body is moved by a linear actuator to up to 152 mm with accuracy of 18 $\mu$m inside the scanner vacuum chamber. The flange-to-flange length of the chamber is 470 mm. The scanner can work in either the vertical or horizontal direction by installing the entire vacuum chamber in the preferred orientation.

One of the design features of the LEBT Allison scanner copied to the MEBT scanner is the sawtooth serrations of the deflection electrodes to prevent particle reflection into the Faraday cup [17]. The serration angle was decreased in accordance with the lower angular range of the beam.

The electronic system supplying the voltages and controlling the scanner motion as well as the control program are shared with the LEBT scanner (described in Ref. [9]). Switching between the two scanners is performed by physically disconnecting one scanner and reconnecting the corresponding cables on the other via a patch panel outside the accelerator enclosure.

### C. Operation

A standard scan with the Allison scanner uses 0.5 mm position steps over 30 mm and 0.5 mrad angle steps over 24 mrad (∼3000 points), taking approximately 5 minutes from the time the scanner intercepts the beam to when it leaves the beam path for a 20 Hz beam pulsing rate. When finer resolution is desired, scans are taken with smaller steps, down to 0.2 mm × 0.2 mrad, typically over a smaller region of interest to reduce the scan time.

The scanner can measure the phase portraits over multiple time bins throughout the pulse, with the smallest size of 1 $\mu$s. This capability can be used to analyze variations of the beam parameters along the pulse. If time resolution is not needed, the bin size is chosen to cover large portions of the pulse to decrease the noise since the Faraday cup signal is integrated over the bin. For a 10 $\mu$s pulse, used in most of studies, the pulse is typically sampled with 5 $\mu$s bins. A bin recorded well after the pulse is subtracted from the data (pixel-by-pixel) before saving the results to decrease instrumental background.

Results of all scans are automatically saved and available for offline analysis.

### D. Background removal

After taking a scan, the scanner operating program, mostly inherited from SNS, removes the background and calculates the RMS parameters of the phase portrait. The background removal is performed by setting to zero all pixels with intensity less than a user-defined threshold. By default, the threshold is set to 1% of the peak intensity, which is adequate to remove the noise for the nominal 5 mA beam. However, for low intensity beams such rejection does not remove all noise, artificially increasing the reported emittance. Also, for high intensity, the cut level can be too aggressive, removing otherwise observable beam tails.

An alternative approach, realized presently only for off-line analysis in a Python [18] script, is to define the threshold based on the level of the noise in a region of the phase portrait where the beam cannot reside and remove only the pixels that cannot be distinguished from the beam signal.

This cut is established in several steps. First, the script finds the area that is most likely to contain only noise. The portrait matrix is divided into four identical rectangles, and the rectangle with minimum total intensity is chosen. A 6×6 pixels square in the outermost corner of this rectangle is assumed to contain only noise signal. The mean signal of this square is then subtracted from each pixel over the entire portrait.

Then, the cut threshold $T_c$ is set proportional to the rms of the noise level $\sigma_n$ calculated in the 6×6 square

$$T_c = A_n \sigma_n \quad (1)$$

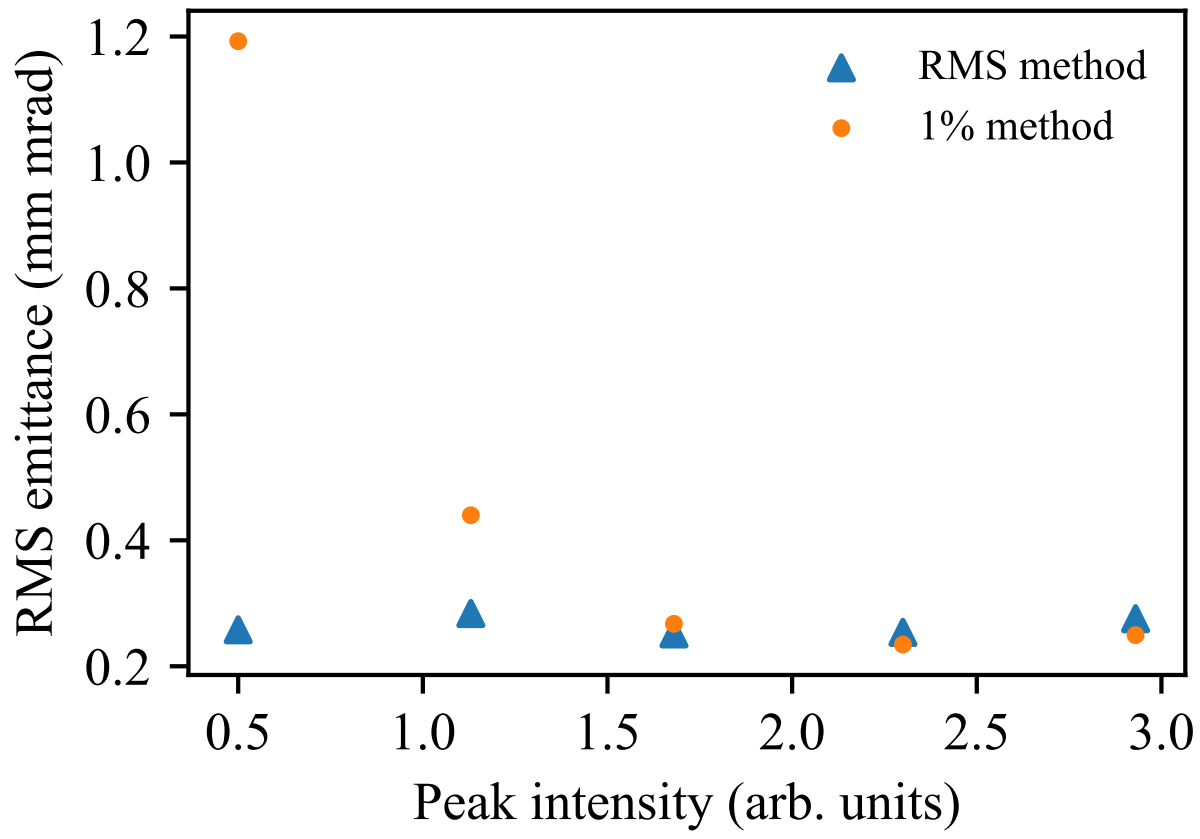

FIG. 4. Vertical RMS emittance with horizontal scraping. The five data points correspond to 1, 2, 3, 4, 5 mA of beam current after scraping.

where the coefficient $A_n$ is chosen according to the procedure described in Appendix A. The intensity of each pixel with initial amplitude below $T_c$ is set to zero.

The real beam is then assumed to be continuous in phase space. Therefore, non-zero intensity pixels with all their neighbors having zero intensity represent the tails of the noise distribution and are removed (i.e. intensity set to zero) as well.

For the nominal beam in PIP2IT, the cut threshold calculated with this method is typically ∼0.5% of the peak intensity.

In this paper, the phase portraits are presented after being cleaned with the described offline procedure, and the reported beam properties derived from these cleaned portraits. For instance, the rms emittance is calculated over all non-zero pixels remaining after cleaning.

As an example of the method benefits, Fig. 4 shows the results of the measurements looking for possible x-y coupling. With the scanner in the vertical position, the horizontal edges of the beam are removed in steps with scrapers. Removal of the beam horizontally results in a lower intensity of a given pixel in the vertical phase space, so that the peak intensity can be used as a measure of the remaining current. When the noise-based cut is used, the measured emittance is constant within 10%, showing that the beam ellipse is not x-y coupled. However, when the same data are analyzed with the 1% cut, the emittance appears to increase when the peak intensity goes below ∼1.5, corresponding to a beam current of roughly 2 mA, due to noise flooding the phase portrait.

## III. PIP2IT BEAM AND BEAMLINE DESCRIPTION

The PIP2IT MEBT in its final configuration is shown in Fig. 5. Transverse focusing is provided by quadrupoles grouped into 2 doublets and 7 triplets; each group includes a Beam Position Monitor (BPM), whose capacitive pickup is bolted to the poles of one of the quadrupoles. The space between each focusing group is referred to as a section. Longitudinal focusing is provided by 3 bunching cavities. One of peculiarities of the MEBT is the bunch-by-bunch chopping system [3], which consists of two travelling-wave kickers separated by ∼ 180° transverse phase advance and an absorber at ∼ 90° phase advance from the last kicker. The absorber is followed by a Differential Pumping Insert (DPI), of which a 200 mm (L) × 10 mm (ID) beam pipe separates the vacuum of the portion of the beam line preceding the future cryomodules from the rest of the warm front end. Movable scrapers [19] are used to measure the beam size, as protection against errant beam or halo, and to intercept one of two trajectories when characterizing the kickers' performance. In the present beam line, there are 4 sets of 4 scrapers (each set consists of a bottom, top, right and left scraper) plus a temporary set of two scrapers (a.k.a. F-scraper, top and right) located just downstream of the prototype absorber before the DPI. For measuring the rms beam size, a scraper is stepped through the beam, and the dump current dependence on the scraper position is fitted to a Gaussian beam profile. The diagnostics set also includes current transformers located at the beginning and end of the MEBT.

TABLE II. PIP2IT MEBT beam parameters

| Parameter | Value | Unit |
|---|---|---|
| Beam energy | 2.1 | MeV |
| Macro-pulse repetition rate | 1-20 | Hz |
| Macro-pulse length | 0.005-25 | ms |
| Bunch repetition rate | 162.5 | MHz |
| Pulse beam current | Up to 10 | mA |
| Transverse emittance, rms norm. | $\leq$ 0.23 | mm mrad |
| Longitudinal emittance, rms norm. | $\leq$ 0.34 | mm mrad |

The beam coming out of the MEBT is absorbed in a beam dump. The main beam parameters are summarized in Table II. The maximum achieved duty factor is 50% with an average power in the dump of 5 kW.

The Allison scanner was used in three locations.

1. In section #1, downstream of the second doublet, in the horizontal position
2. In section #5, in lieu of the absorber prototype, in the vertical position
3. Toward the end of the line as shown in Fig. 5, in the vertical position.

## IV. CALIBRATION AND PERFORMANCE OF THE ALLISON SCANNER

Calibration of the position coordinate was performed using a micrometer attached to the emittance scanner body and recording the number of counts of the stepper motor for a given displacement. Measurements of the position showed excellent linearity over the range of motion required for typical measurements (30 mm), while accuracy of the displacement was determined to be ±18 $\mu$m (rms) for displacement requests ranging from 0.25 mm to 5 mm, independently of the direction (inward or outward).

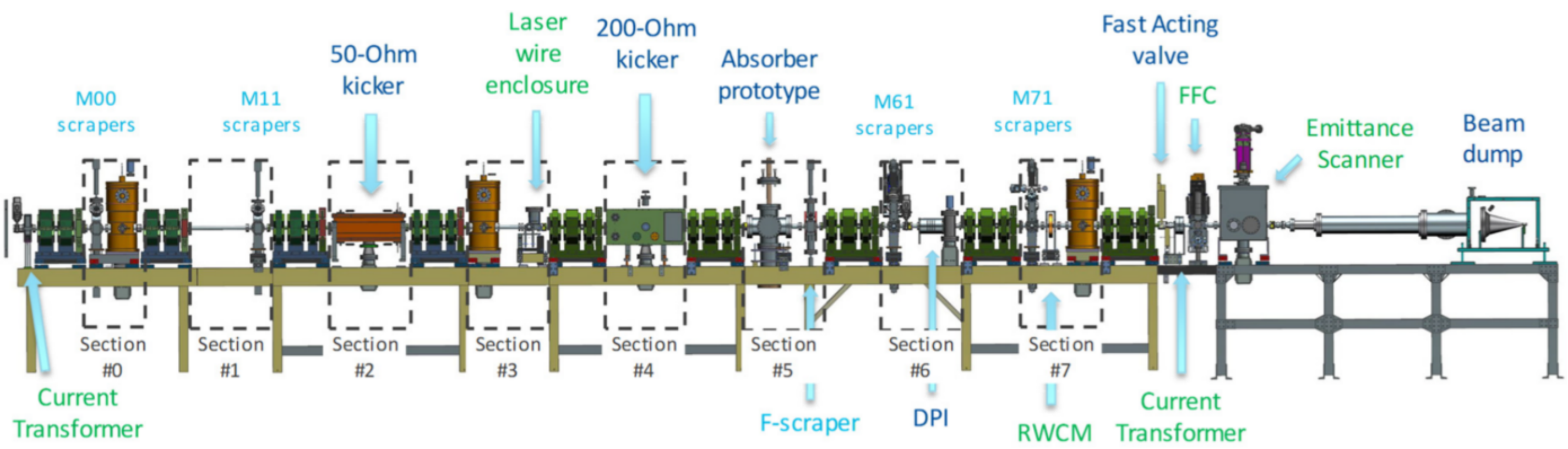

FIG. 5. Medium energy beam transport line (side view).

The initial calibration of the angle coordinate for a given voltage was determined by simulations of the geometry (electrostatic calculation followed by single-particle tracking). The calibration was checked with the beam. The procedure, described in detail in Ref. [9] for the LEBT Allison scanner, consists of steering the beam with a corrector upstream of the emittance scanner and recording the beam's centroids displacements in position and angle. In free space, they are proportional, and the corresponding coefficient is the distance between the corrector and the front slit of the emittance scanner, which is known to a good accuracy.

Calibration of the scanner was carried out in a configuration with a short MEBT (Fig. 6), composed of 2 quadrupole doublets with BPMs, 2 scraper assemblies, 1 bunching cavity upstream of the scanner vacuum chamber, and the scanner oriented horizontally. To carry out the calibration, the second doublet and the bunching cavity were turned off in order to create a drift space between the dipole corrector magnet after the first doublet and the Allison scanner.

Figure 7 shows the centroid angle versus the beam centroid position as recorded by the emittance scanner. The inverse of the slope on Fig. 7 is the distance between the origin of the kick (steering corrector) and the emittance scanner entrance slit and is 953.5 mm (with a standard error of 2 mm rms), which differs from measurements made with laser tracking by less than 0.3%. Therefore, the initial calibration of the scanner angle was not changed. Cross-checks of the angular calibration with BPMs and scraper data also agree well with expectations.

Note that the rms emittance is significantly more volatile, with measurement-to-measurement variation up to 5%. While part of this variation is related to the electronics noise, likely it also reflects an actual variability of the beam phase space.

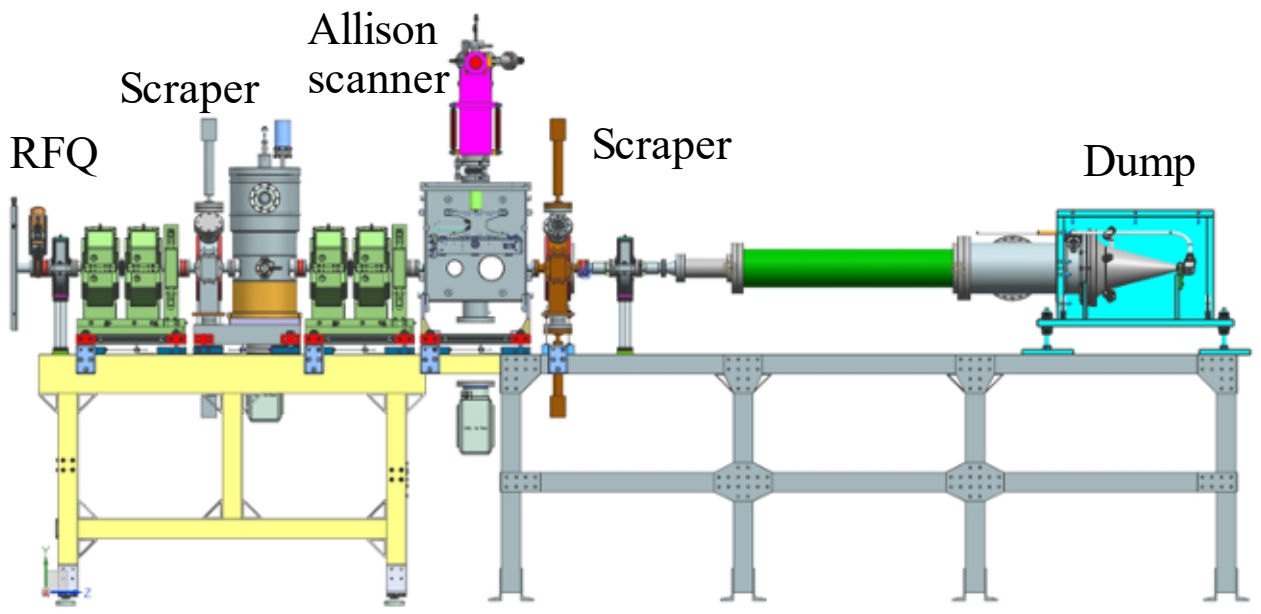

FIG. 6. MEBT configuration for calibration of the emittance scanner

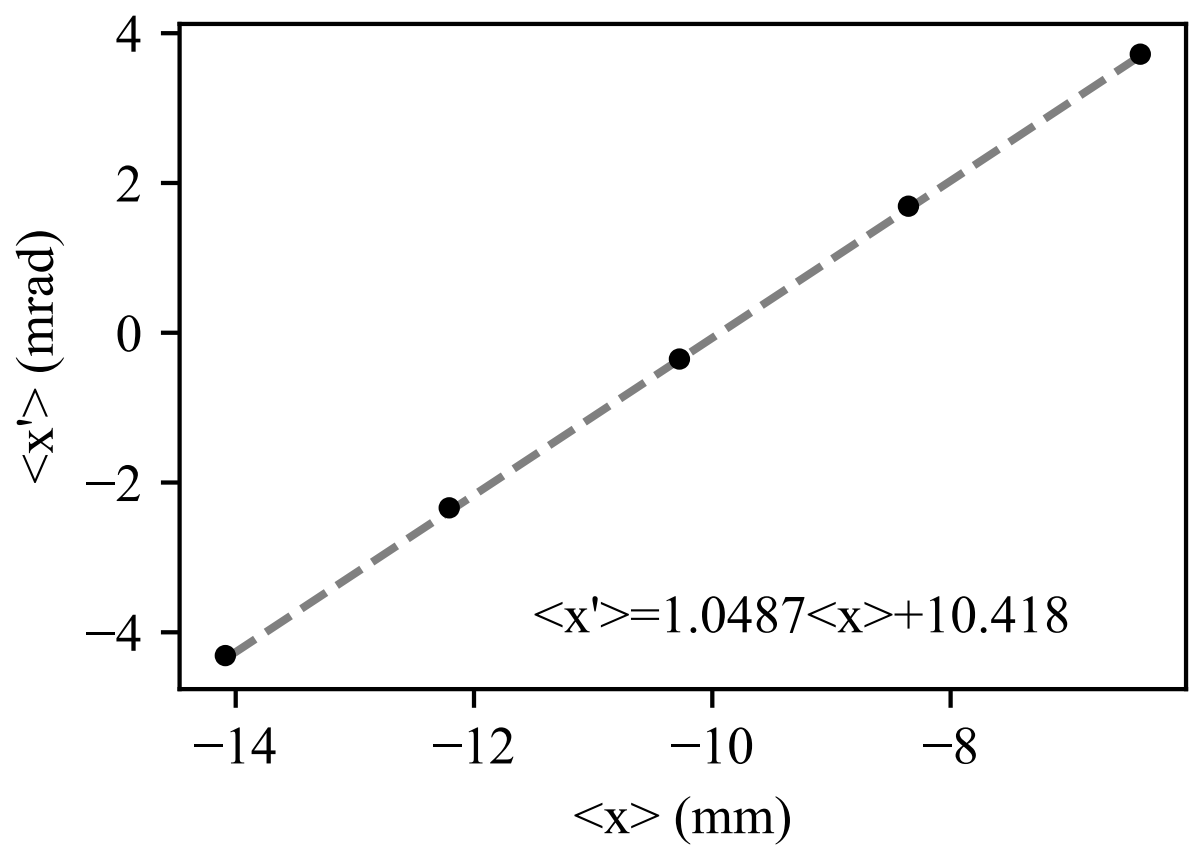

FIG. 7. Angular calibration of the Allison scanner.

The rms emittance measured with the Allison scanner was compared to the results from a quadrupole scan in the configuration shown on Fig. 6. The quadrupole scan was performed by varying the strength of the most downstream quadrupole and measuring the rms beam size with the horizontal and vertical scrapers. This scan measured an rms emittance of 0.17 mm mrad which is slightly below the average value measured with the Allison scanner.

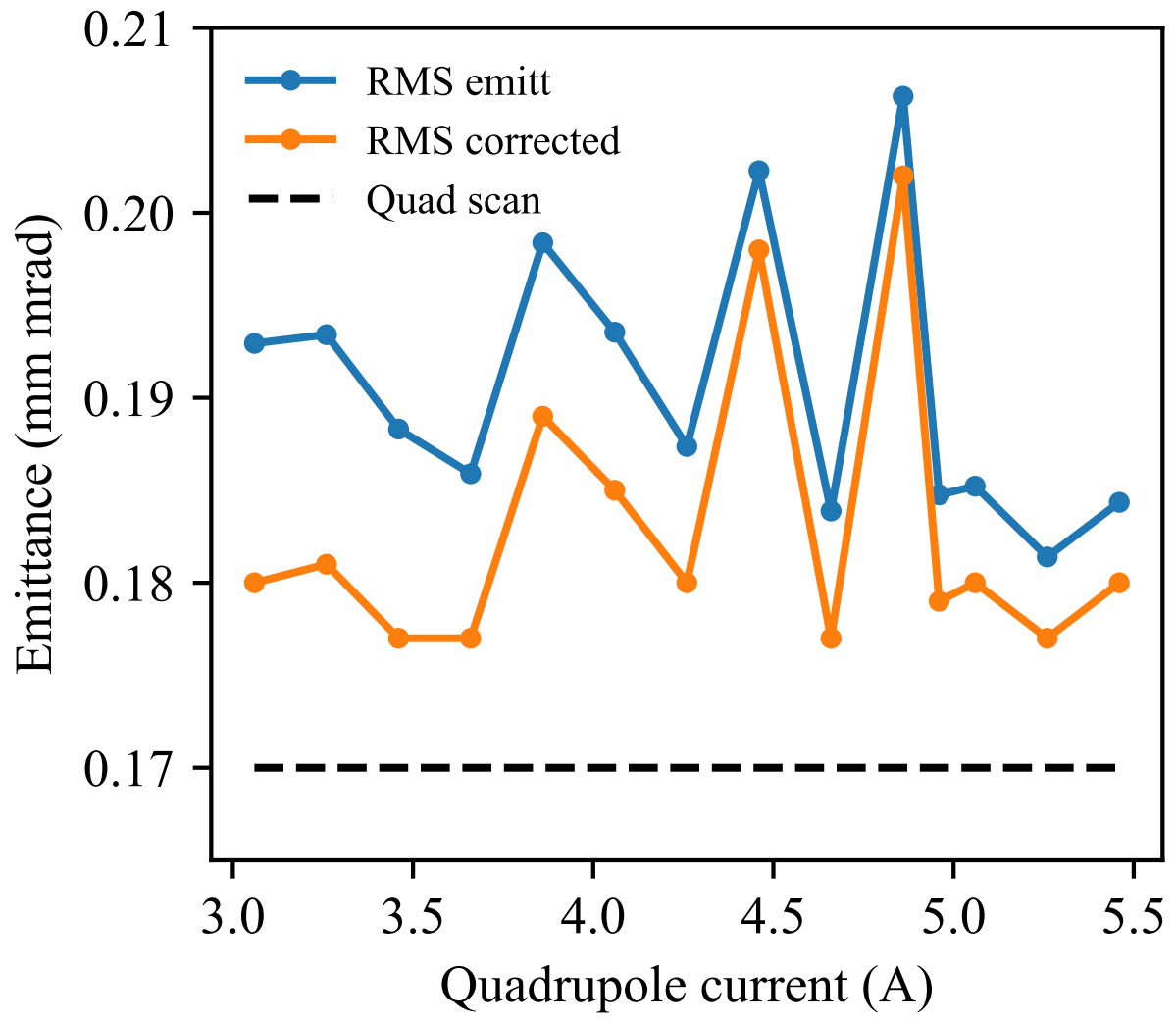

FIG. 8. RMS emittance measured with the Allison scanner with and without correction for the slit size and emittance measured from a quadrupole scan.

For the comparison, one needs to take into account that the values of the rms parameters directly calculated from the phase portraits are affected by the finite size of the scanner slits. In approximation of a 2D Gaussian distribution, this effect can be estimated and corrected according to formulae in Ref. [20]. For the rms emittance, typical values of the correction are 2-7% depending on the Twiss parameters (Fig. 8). After correcting for the effect of the slit size, the emittance measured with the Allison scanner is constant to within ±5% with the average value being approximately 10% larger than the quadrupole scan result for the horizontal emittance.

Another effect increasing the reported emittance is related to the finite measurement time in the presence of a beam jitter. The beam centroid was found to be moving significantly in the PIP2IT MEBT, with rms amplitude of up to 0.2 mm, depending on location and optics setting [5].

In location (1) of the Allison scanner (as specified in Section III) the motion has a negligible effect on the phase portraits, mainly because the scanner measured the horizontal plane while the jitter is predominantly in the vertical plane. In the other two locations, the scanner measured the vertical phase space of the beam and the effect is visually noticeable resulting in less smooth phase portraits (e.g. Fig. 11(a)).

The beam motion appears to be random in time with frequencies up to ∼3 Hz. Therefore, even an individual angular scan is significantly affected (∼20 meaningful points are measured across the beam at 20 Hz results in ∼1 s scan time). In the limit-case model when the beam centroid shifts are completely uncorrelated during recording of any pixel and the number of pixels is large, the scanner measurement represents the time-average emittance, i.e. the average area occupied by beam in phase space over a long time. While the time-average emittance, rather than the emittance of the beam distribution, is the one that matters for the beam transport, the inability to measure an "instantaneous" phase portrait complicates the analysis of beam tails. However, the beam centroid motion is still much lower than typical beam dimensions in the phase space, and the changes in the rms values are small. An independent estimate using BPM data indicates that the time-average emittance exceeds the "instantaneous" emittance only by ∼2%, less than the statistical scatter of the measurement. Hence, in this paper, we will not distinguish between them.

## V. PARTICLE DISTRIBUTION PARAMETERS

### A. Quantitative comparison of phase portraits

One of the advantages of using the Allison scanner is the ability to measure the 2D phase portrait of the beam as opposed to only measuring its rms values. This allows for analyzing in detail the phase density distribution, including the beam tails. However, understanding the dynamics of the distribution from direct comparison of different phase portraits in x-x' (position – angle) coordinates is complicated since the portraits may differ dramatically when the Twiss parameters change even for purely linear optics [21]. For quantitative analysis, we found more appropriate to describe the phase portraits in action-phase coordinates. The action $J$ and phase $\phi$ are defined as

$$J = \frac{1}{2}\left(\gamma x^2 + 2\alpha x x' + \beta x'^2\right) \tag{2}$$

$$\phi = \arctan\left(\frac{\alpha x + \beta x'}{x}\right) \tag{3}$$

where $\alpha$, $\beta$, and $\gamma$ are the Twiss parameters and $x$ and $x'$ are the position and angle coordinates of a particle [22].

Note that for convenience of comparison with the normalized rms emittance, the action plotted in this paper is normalized, i.e. multiplied by the product of the relativistic factors $\beta_r\gamma_r = 0.0668$.

In linear optics with no x-y coupling, the action is a constant of motion, i.e. the action of a particle remains the same for different optics settings. Therefore, a description of the particle distribution over the action provides a more stable description of the beam, allowing for easier distinction between possible instrumentation artifacts and actual changes in the beam properties.

### B. Core description

Analyses of the phase portraits measured in the MEBT in action-phase coordinates, revealed that in the central portion of the beam, i.e. at low actions, the pixel intensity is largely independent from the value of the phase and decreases exponentially with the action. A simple fit thus follows

$$I_{\text{gauss}} = I_0 e^{-J/\epsilon_c}. \tag{4}$$

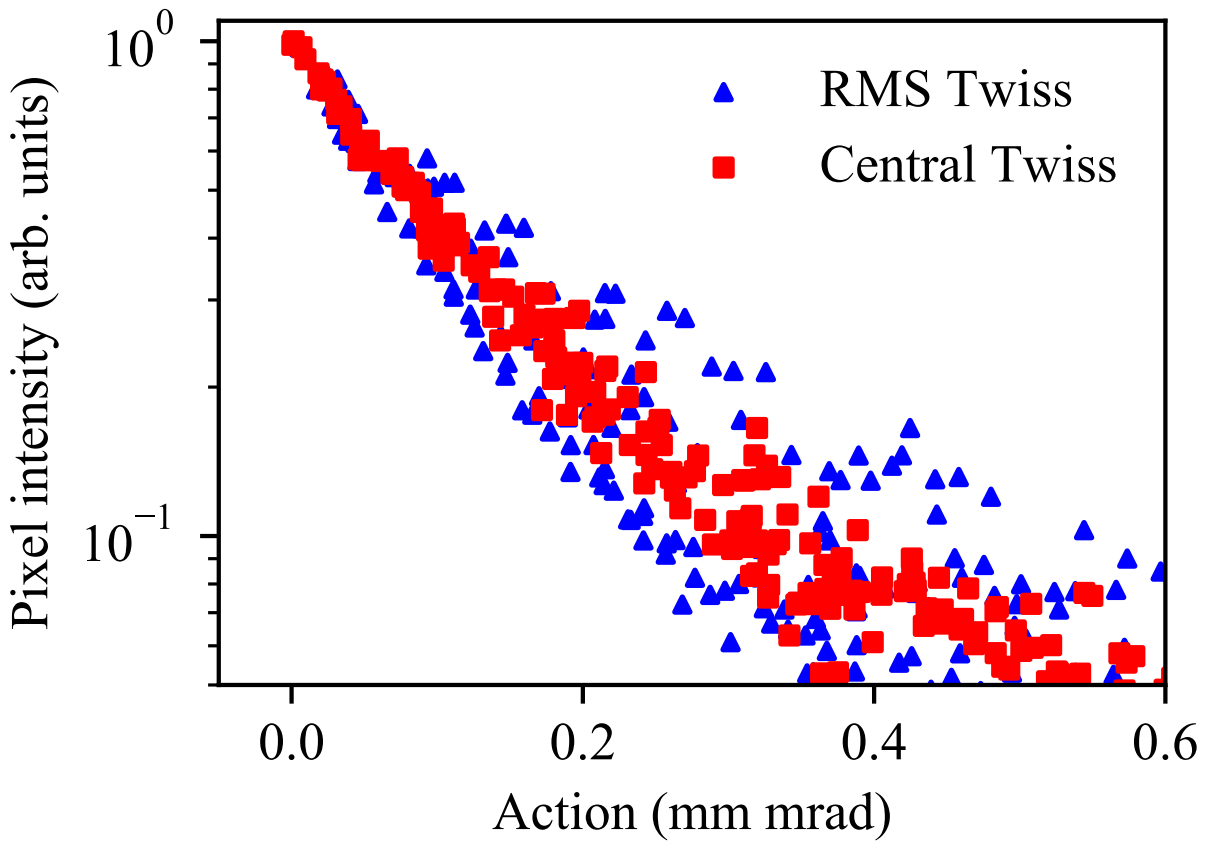

FIG. 9. Action distribution using central Twiss parameters (red) and rms Twiss parameters (blue). Using the RMS parameters to define the emittance results in a wide range of intensities for a given action even close to the center of the beam. Using central Twiss parameters narrows the distribution at low action as expected from Eq. (4).

In semi-logarithmic coordinates, customary for presenting the beam distribution over action, Eq. (4) is a straight-line which slope is given by $-1/\epsilon_c$. The parameter $\epsilon_c$ will be referred to in this paper as the "central" slope. Since Eq. (4) describes a perfect Gaussian distribution, the central slope can be interpreted as the rms emittance of the beam if the Gaussian core were extended and the tails removed.

The first attempts to compare the measured data with Eq. (4) showed a relatively large scatter of pixel intensities for any given action, even for the central part of the beam (Fig. 9, blue). We put significant efforts to identify possible procedural and instrumental peculiarities that can cause such scatter.

One consideration is the dependence of the distribution over action on the choice of the Twiss parameters used to define the action. For the distribution of Fig. 9, blue, the Twiss parameters are the rms parameters of the entire beam (referred here as "rms Twiss parameters"). This choice of Twiss parameters results in a large scatter even for particles with low action.

Alternatively, the calculation can be done by directly fitting Eq. (4) to the measured distribution with the Twiss parameters, central slope, and the peak intensity ($I_0$ in Eq. (4)) as free parameters. The resulting fit depends on the choice of the ensemble. When fitting, the distribution is cut in intensity to remove the tails so they do not affect the definition of action. If the cut is too small, the tails cause the central slope to increase. However, it was found that if the cut is too large, causing the number of remaining pixels to be below ~30, the small number of remaining pixels causes large statistical uncertainty in fitted Twiss parameters. The procedure for determining the size of the cut and defining the Twiss parameters from the central region is described in Ref. [23].

When these so-called "central Twiss parameters" are used, the scatter in the beam's central region is decreased significantly (Fig. 9, red). One of the ways to quantify this decrease is to compare the standard error of $\epsilon_c$ found from fitting the upper 50% of the intensity of either distribution to Eq. (6) in Ref. [23]. The rms and central Twiss parameters yield, correspondingly, 0.155±0.061 and 0.109±0.014 mm mrad, i.e. the error is reduced by a factor of 4 when using the central Twiss parameters.

Note that the finite size of the scanner slits not only affects the rms measurements but also modifies the the details of the distribution. As it is shown in Ref. [23], distribution of a purely Gaussian beam reconstructed with the Allison scanner appears to be phase – dependent, with the amplitude of the artificial $\cos(2\phi)$ component determined by the beam parameters. The formula of Ref. [23] can be applied for correcting the measured distributions. Such procedure is not quite self-consistent since the formula is based on a Gaussian beam distribution while the measured distributions are more complex. However, it does correct most of the effect. In part, reconstruction of the central slope become significantly more stable (see Fig. 2 in Ref. [23]). On the other hand, modifications for individual pixels intensities are minor. As a result, the slits correction is only used for fitting the central parameters, while all plots show intensities that have not been corrected.

### C. Discussion on the beam distribution

Note that while the assumption about the beam following Eq. (4) is common in the description of beam dynamics, it was not obvious for us that it would adequately describe the portraits measured at the PIP2IT MEBT. One of the considerations was that the distribution of the beam coming out of the ion source significantly deviates from Eq. (4) for all actions since the beam is initially spatially limited by the ion source extraction aperture. This feature is clearly seen in the portraits recorded near the ion source and in the LEBT and was beneficially used in the design of the partially un-neutralized transport scheme for the LEBT [15]. The LEBT phase portraits can be approximated by distributions initially Gaussian in velocities and constant density in the radial direction, referred as Uniform-Gaussian (UG) in Ref. [15]. Eq. (7) there, describing the UG distribution, projected into one plane and expressed in terms of action-phase, is very

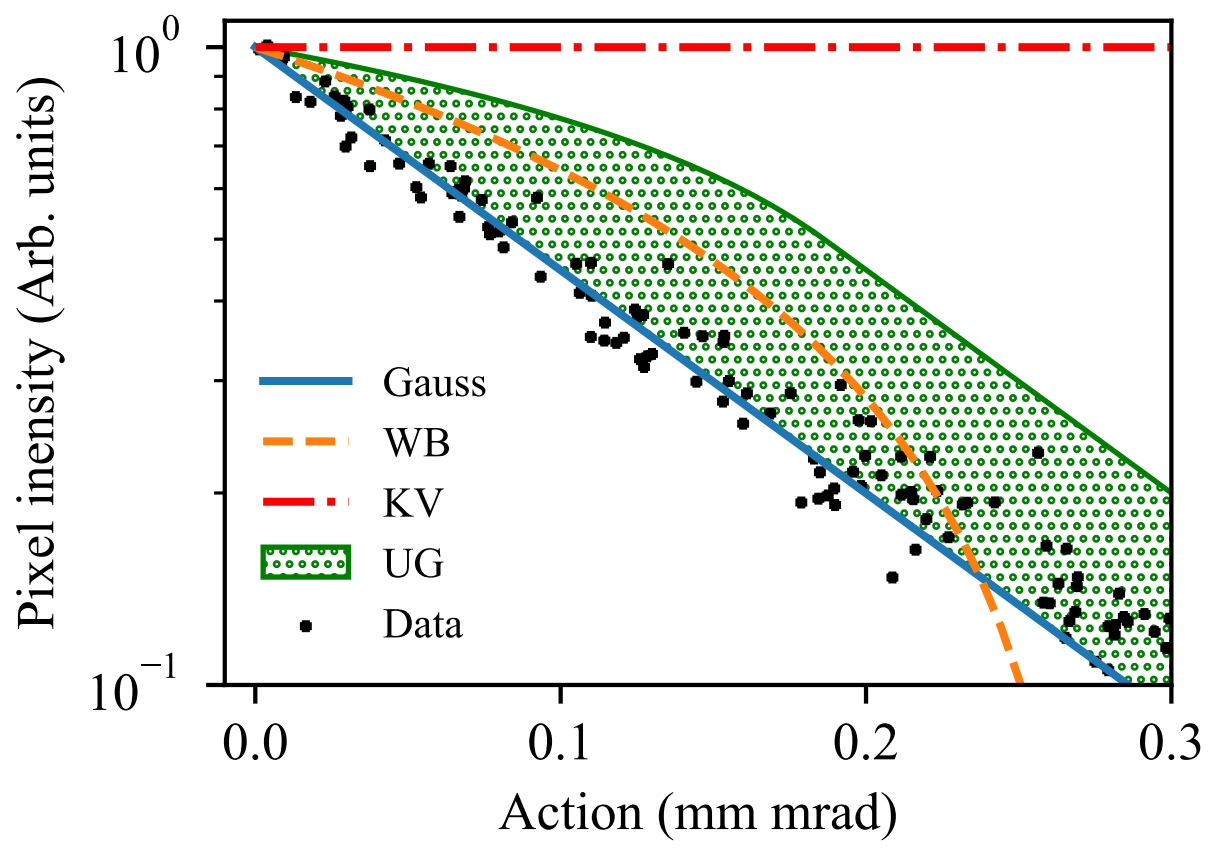

FIG. 10. Comparison of the measured distribution in action in the beam core (black) with several idealized distributions: Gaussian, KV, UG, and WB. Note that the UG distribution is phase-dependent, and, therefore, pixel intensities vary for a given action.

different from the Gaussian's:

$$I_{UG}(J,\phi) = \frac{I_0}{\pi\sqrt{2\pi}\epsilon_{UG}}\sqrt{1-\frac{J\cos^2(\phi)}{2\epsilon_{UG}}}e^{-\frac{J\sin^2(\phi)}{\epsilon_{UG}}} \times \begin{cases} 1 \text{ if } \frac{J\cos^2(\phi)}{2\epsilon_{UG}} \leq 1 \\ 0 \text{ else} \end{cases} \quad (5)$$

where $\epsilon_{UG}$ is the rms emittance.

Kapchinskiy-Vladimirskiy (KV) and waterbag (WB) distributions are also commonly used, and they also have significantly different forms than Eq. (4) in action-phase coordinates.

$$I_{KV}(J,\phi) = I_0 \quad (6)$$

$$I_{WB}(J,\phi) = I_0\left(1-\frac{J}{2\epsilon_{WB}}\right) \quad (7)$$

According to Ref. [24], after experiencing multiple betatron oscillations, the beam is expected to relax toward the Maxwell-Boltzmann distribution, which, for the parameters of the PIP2IT MEBT, corresponds to Eq. (4). Our conclusion from fitting the data to different idealized distributions is that the ~12 betatron periods going through the RFQ are sufficient to fully achieve this relaxation in the beam core. It is illustrated by Fig. 10, where the high-intensity pixels, containing 85% of the beam, are plotted against the action together with several idealized distributions. Clearly, the fit with Eq. (4) best describes the data.

### D. Tail description

It is customary to describe the beam as a composition of two or three parts [21, 25], e.g. core and tails, however, there is a lack of uniformity in the definitions, e.g. [26, 27], with each one having its own merits. For the purpose of this paper, we suggest separating the core and tails according to the pixels' actions. The procedure to define the "transition action" $J_{tr}$ to separate the tails from the core is chosen as follows.

First, the central parameters are determined as outlined above and the action and phase are calculated for each pixel. Then, all the pixels are sorted into action bins $J_i$, typically 0.05 mm mrad in size, and the mean intensity $I(J_i)$ and the standard deviation $\sigma_{Int}(J_i)$ of the intensity in each action bin is calculated. The value of $J_{tr}$ is defined as the action of the bin where the mean intensity deviates from the fit of Eq. (4) by more than three times the standard deviation of the mean,

$$I(J_{tr}) - I_0 e^{-J_{tr}/\epsilon_c} = 3\sigma_{Int}(J_{tr}) \quad (8)$$

All particles with action less than the transition action are defined to be in the core, and particles with larger action are in the tail. The percentage of the beam in the tails is typically about 10-20% of the total intensity.

With this definition, the transition action and percent of the beam in the core are constant under linear optics (in practice, within the statistical noise). Therefore, these two parameters can be used as a metric for tail growth due to non-linear effects.

At actions above $J_{tr}$ the scatter of pixel intensities at a given action visibly increases and clearly deviates from the Gaussian core. The dominant part of this scatter comes from strong phase dependence with the tail being split into two "branches" of similar intensities that are separated in phase by approximately $\pi$ rad. This observation is even more evident when the data is plotted in $J-\phi$ coordinates (Fig. 11). The location in phase of the branches is defined by averaging over pixels with $J > 1.5J_{tr}$ and phases centered around the peak within $(0,\pi]$.

Hence, in this paper, the measured beam distribution is described, in addition to the traditional rms values, by seven parameters. The beam core is characterized by the central slope $\epsilon_c$ and central Twiss parameters $\alpha_c$, $\beta_c$. Pixel intensity there decays exponentially with action and is independent on the phase. The intensities eventually deviate from this behavior at the transition action $J_{tr}$. All particles with action larger than the transition action are in the tails which are characterized by the phase of the branches $\phi_b$, the maximum action $J_{\max}$, and the fraction of the particles in the core. Unfortunately, attempts to find an analytical description of the tail distribution did not succeed. Also, so far, we have not been able to reliably identify the origins of the tails.

Obviously, a portion of the beam below the noise threshold is not represented in this analysis. A convincing way to estimate what percentage of the actual beam is presented in the recorded phase portraits was not found. However, by extrapolating the curves in Fig. 13(c) to infinity, we estimate that it exceeds 99%.

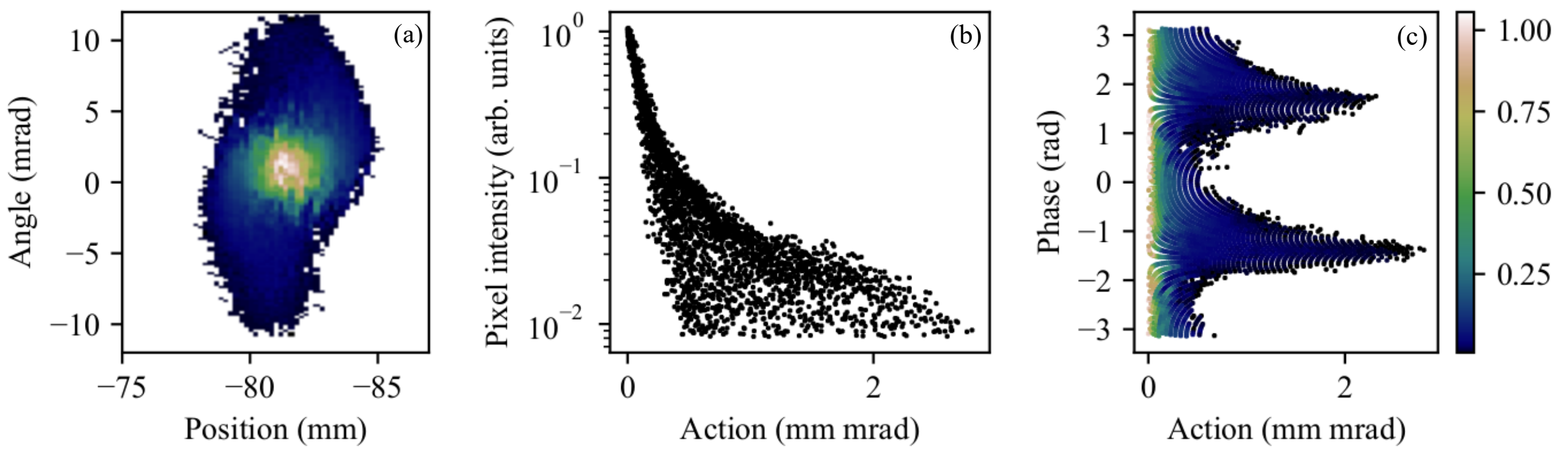

FIG. 11. Phase portrait in position-angle phase space and action-phase phase space. The beam splits into two branches separated in phase at large actions.

## VI. SELECTED BEAM MEASUREMENTS

The Allison scanner was widely used in the PIP2IT MEBT for beam characterization and tuning. In this section, we present several significant results that highlight the use of action-phase coordinates and the usefulness of the Allison scanner for commissioning of the MEBT.

### A. Quadrupole scan

One of the concerns with recording phase portraits is the possibility of the data being contaminated by secondary particles. The most straightforward way to identify such contribution is to compare the distributions recorded with different optics settings and verify that they change according to expectations for the primary beam. The simplest test is to scan a quadrupole magnet close to the Allison scanner so no significant changes in the distribution over action are expected. Results of a such scan are presented in Fig. 12, where the data used to obtain Fig. 8 are further analyzed. The portion of particles in the core and the central slope are found to be stable (Fig. 12(b)) within ±3% and ±5%, correspondingly. Despite the dramatic visible changes of the portraits in x-x' coordinates (Fig. 12(a)), the distribution in action-phase coordinates stays the same (Fig. 12 (d,e)), and portion of particles outside of a given action is stable for more than 99% of the beam (Fig. 12 (f)).

Note that in a given portrait the particle phase is defined according to Eq. (3) with respect to the particles with zero canonical angle, i.e.

$$\phi = 0 \text{ at } x'_c \equiv \frac{\alpha x}{\sqrt{\beta}} + x'\sqrt{\beta} = 0. \tag{9}$$

Therefore, the phase of the particles for scans with different optics shifts by the difference in betatron phase advance between these portraits. While a phase shift cannot modify the appearance of the phase-independent core, the phase position of the tails should change accordingly. In the case of the presented quadrupole scan (as well as in other quadrupole scans recorded), the actual change in the phase advance is small because the distance between the varied quadrupole and the Allison scanner is small. The observed unchanged phase position of the branches, within measurement errors, is in agreement with the simulated phase advance (Fig. 12 (c)).

The emittance measured with the quadrupole scan, 0.17 mm mrad (Fig. 8), is between the rms (0.18 mm mrad) and the central slope (0.14 mm mrad) values measured with the Allison scanner. It may be related to the procedure of fitting the scraper measurements to Gaussian distributions used in the quadrupole scans, which is predominately sensitive to the beam core.

In summary, the quadrupole scan analysis shows a stable distribution in action and did not show any contribution of secondary particles.

### B. Comparison of measurements in different locations

The stability of description of the distributions in action-phase coordinates allows for comparisons of the phase portraits of the beam that have passed through significantly different optics. As it has been mentioned in Section III, phase portraits were recorded in three locations and with two scanner orientations over 18 months. The results of measurements, performed with the same settings for the ion source, LEBT, and RFQ, are summarized in Table III. Each result presents an average over 10 measurements made on different days in an attempt to separate day-to-day variability from difference between locations and orientation. Errors are the RMS error over each set of 10 measurements.

Across all three locations of the Allison scanner, the RMS emittance is the same within these errors. Also, no change, within the scatter, is observed in the central

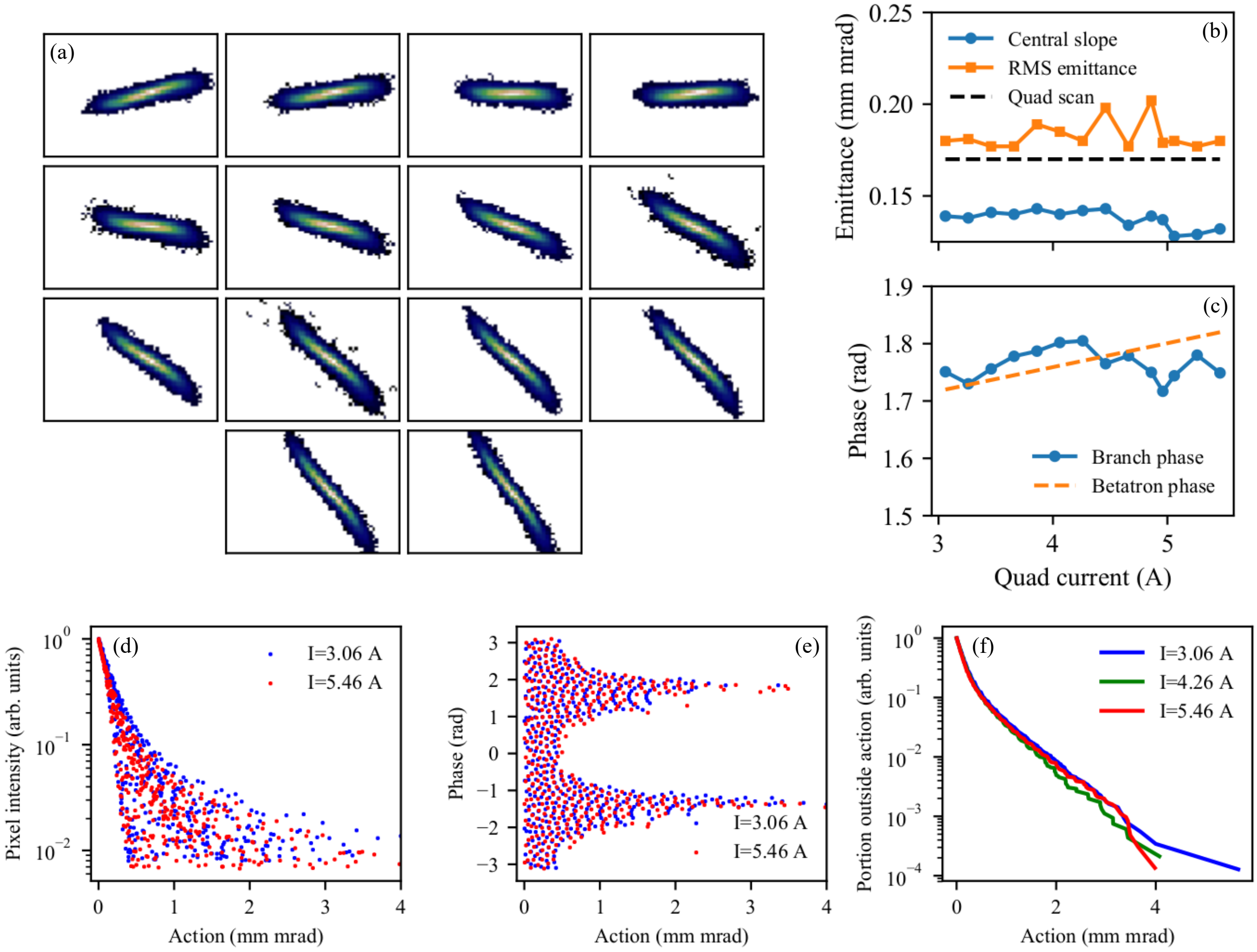

FIG. 12. Analysis of a quadrupole scan. (a) phase portraits in (x,x') coordinates recorded at the quadrupole currents increasing from left to right and from top to bottom from 3.06A to 5.46 A. The x and x' ranges in each plot are 30 mm and 24 mrad, correspondingly. No significant variation of the slit - corrected central slope and percent in the core are observed while a quadrupole strength was scanned (b). The average branch phase and betatron phase also did not change (c). Phase portraits in action-phase coordinates for the minimum and maximum quadrupole currents overlap (d), (e). The portion of the beam outside of a given action is stable over most of the beam (f).

slope and fraction of intensity in the core from location 2 to location 3, in which the Allison scanner measured in the vertical plane. We interpret this as an absence of measurable changes in the beam core parameters in the MEBT. Higher values of the central slope and percent in the core at the location 1 are attributed to the difference between the horizontal and vertical planes, since these values stay constant from location 2 to location 3. Although, direct comparison by measuring both planes in single location was not performed. Outside of 99% of the measured beam intensity, the difference between distributions is larger that one would expect from statistical fluctuations and reconstruction errors by comparing with Fig. 12 (f). The increase of particle population outside of large actions from location 2 to location 3 visible in

TABLE III. Average rms emittances and core and tail parameters for the three locations of the Allison scanner. The beam current is 5 mA.

| Location | rms $\epsilon$ | $\epsilon_c$ | % in core |
|---|---|---|---|
| 1 - horz | $0.20 \pm 0.013$ | $0.146 \pm 0.003$ | $88 \pm 2.5$ |
| 2 - vert | $0.19 \pm 0.015$ | $0.117 \pm 0.013$ | $71 \pm 11$ |
| 3 - vert | $0.22 \pm 0.024$ | $0.123 \pm 0.011$ | $72 \pm 10$ |

Fig. 13 (c) might be interpreted as a halo growth.

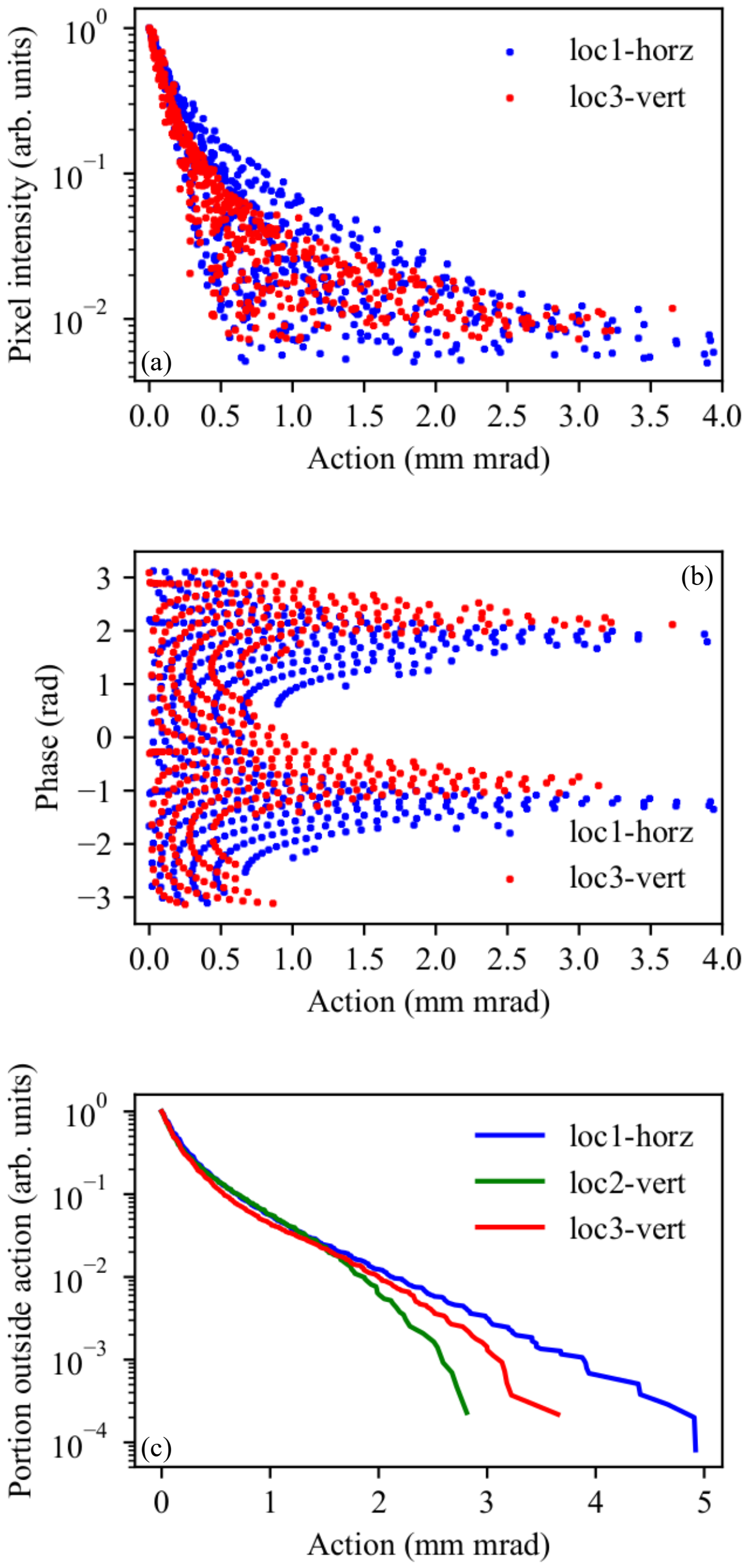

FIG. 13. Comparison of the action distributions at the beginning and end of the MEBT. Note the differences are attributed to the change in measurement plane not propagation along the MEBT.

### C. Beam evolution through the pulse

Most of measurements in the MEBT are performed with 10 $\mu$s pulses to minimize possible damage to the beam line and to insertable diagnostics. On the other hand, the beam parameters interesting for operation are in a steady state or at least in sub-ms range. To understand the variation of the beam parameters during the pulse, several sets of measurements with 0.5 ms pulse duration are performed with the Allison scanner, recording the data in 10 $\mu$s bins. To avoid damaging the front slits, the current density is reduced to a safe level by increasing the beam size to 3 mm rms in each direction.

With the nominal LEBT settings (Fig. 14 red curves), all the parameters are constant within about $\pm 5\%$ rms across the pulse. This allows to study the MEBT (and, in the future, the SRF components) with short pulses, which are less likely to damage the machine than nominally long pulses. Note that the beam parameters where found to be significantly dependent on the settings of the LEBT electrodes controlling the beam neutralization. As an example, the blue curves in Fig. 14 show beam parameters for the case of the ion clearing voltage turned off.

### D. Distribution at different beam currents

Multiple studies were performed to optimize operation by adjusting the ion source settings. As an illustration, Fig. 15 presents the measurements of the phase portraits of a 20 $\mu$s pulse beam with the Allison scanner in location 1 while varying only the ion source extraction electrode $V_{extr}$ to change the beam current. All other settings, tuned to optimize performance at 5 mA, are kept constant.

Nominally the extraction voltage was kept around 3 kV. Decreasing $V_{extr}$ below 3 kV results in a dramatic drop in the RFQ transmission (deviation from the straight line in Fig. 15(a)) and in a growth of the rms emittance (Fig. 15(c)) even near 5 mA operation.

On the other hand, increasing Vextr further above 3 kV decreases the beam current without dramatically changing the beam quality or Twiss parameters (Fig. 15(d)). This allowed for measurements at different beam intensities (in the range of 2 -5 mA for these specific settings) by changing only the extraction voltage, without re-tuning the entire beam line.

For higher beam currents, the the beam size in phase space is larger as seen in the RMS emittance and the central slope. However, the net result is the higher peak phase space density, visible in the phase portraits as increase of the maximum pixel intensity for higher beam currents. Therefore, to generate a bright, low-emittance beam for initial tuning it is preferable to heavily scrape a 5 mA beam in the MEBT to keep the higher intensity core rather than decreasing the beam intensity from the ion source.

### E. Scraping

The scraping system in the PIP2IT MEBT is designed to protect the machine from errant beam and halo. When the scrapers in each location are moved to the beam boundary, it limits the maximum action of particles coming through, protecting the downstream elements. Intercepting part of the halo with the scraping system was foreseen as a normal mode of operation, with preliminary estimates made for a phase-independent Gaussian beam in Ref. [19]. The situation in the experiment was found to be significantly more complicated due to the phase-dependent branches, and the phase portraits recorded with the Allison scanner for different scraping scenarios were highly instructive. For illustration, Fig. 16 compares phase portraits recorded when removing beam with a single scraper at different locations. For this study the top scraper was moved into the beam at each of the stations, one at a time, to intercept 10% of the current (0.5 mA) based on the measured current at the beam dump. In Fig. 16 phase portraits with (red) and without (blue) scraping are overlapped in x-x' (left) and $J-\phi$ (right) coordinates. The action and phase of the scraped beams are calculated using the center of charge and the central Twiss parameters of the non-scraped beam.

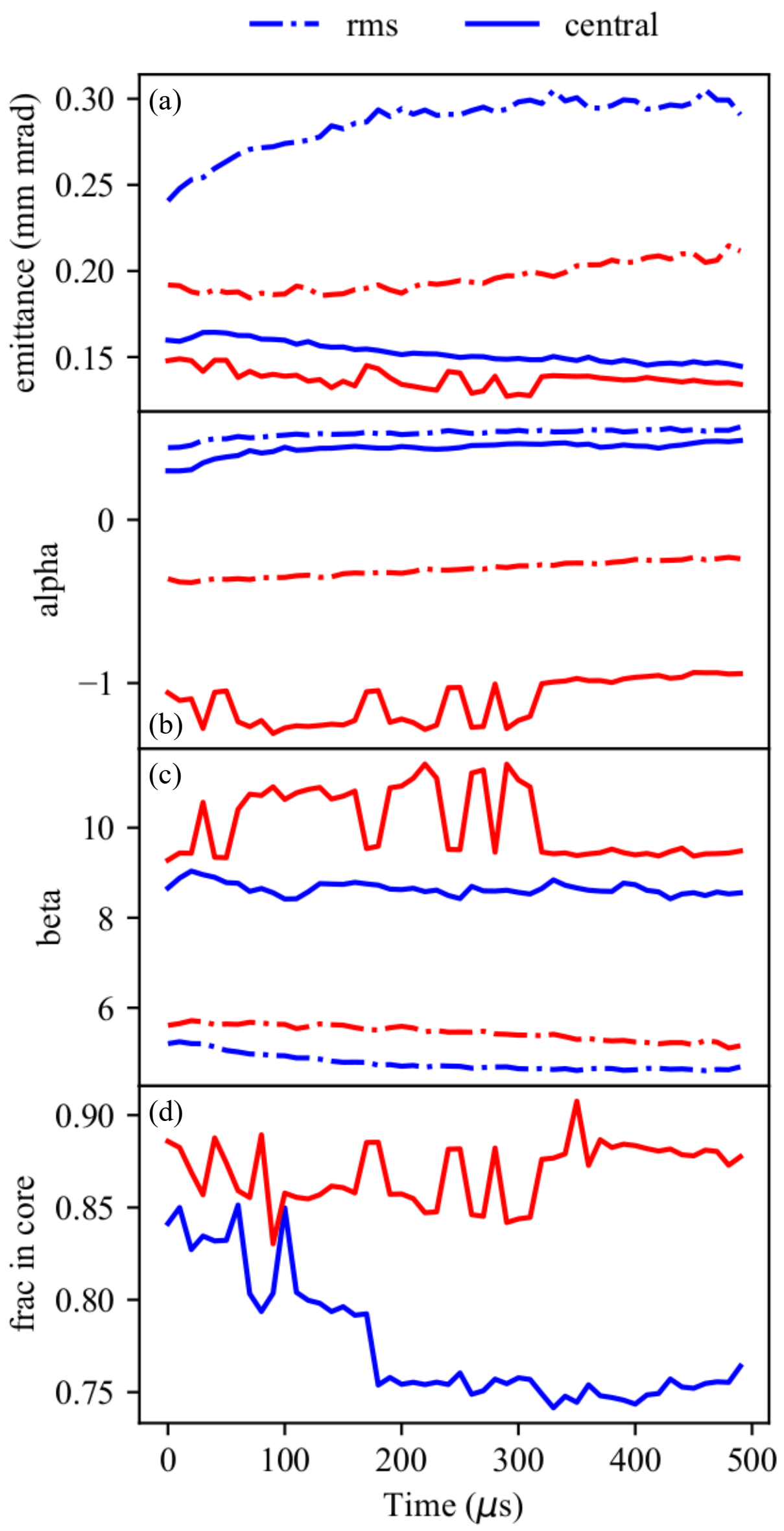

FIG. 14. Parameters evolution through a 0.5 ms pulse for nominal LEBT settings (red) and perturbed settings (blue). In the latter case, the ion-clearing voltage in the LEBT is decreased from 300 V to 0 V.

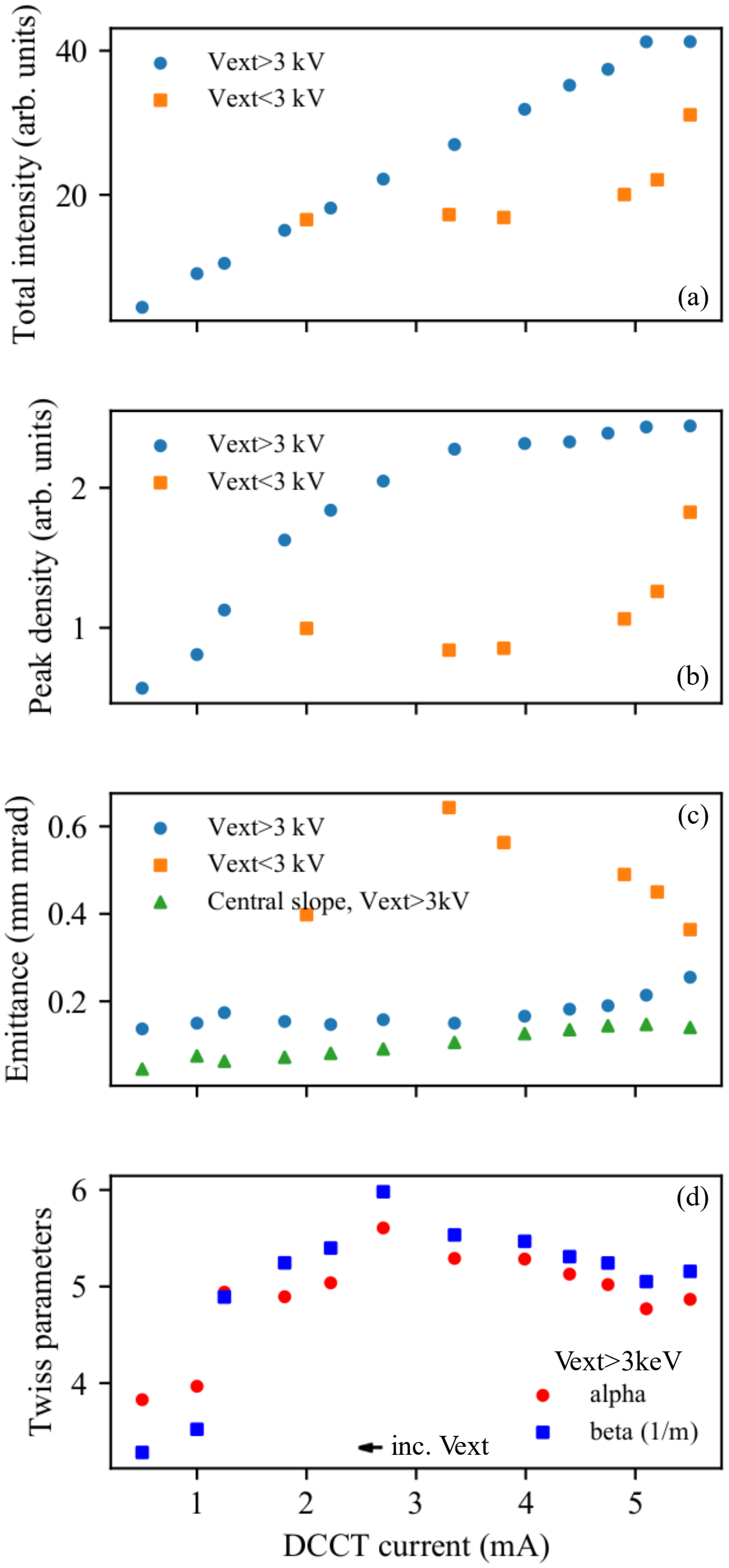

FIG. 15. Beam parameters for different extraction voltages $V_{extr}$. Parameters are plotted as functions of the beam current in the LEBT.

Figure 16 shows that scraping the same fraction of the

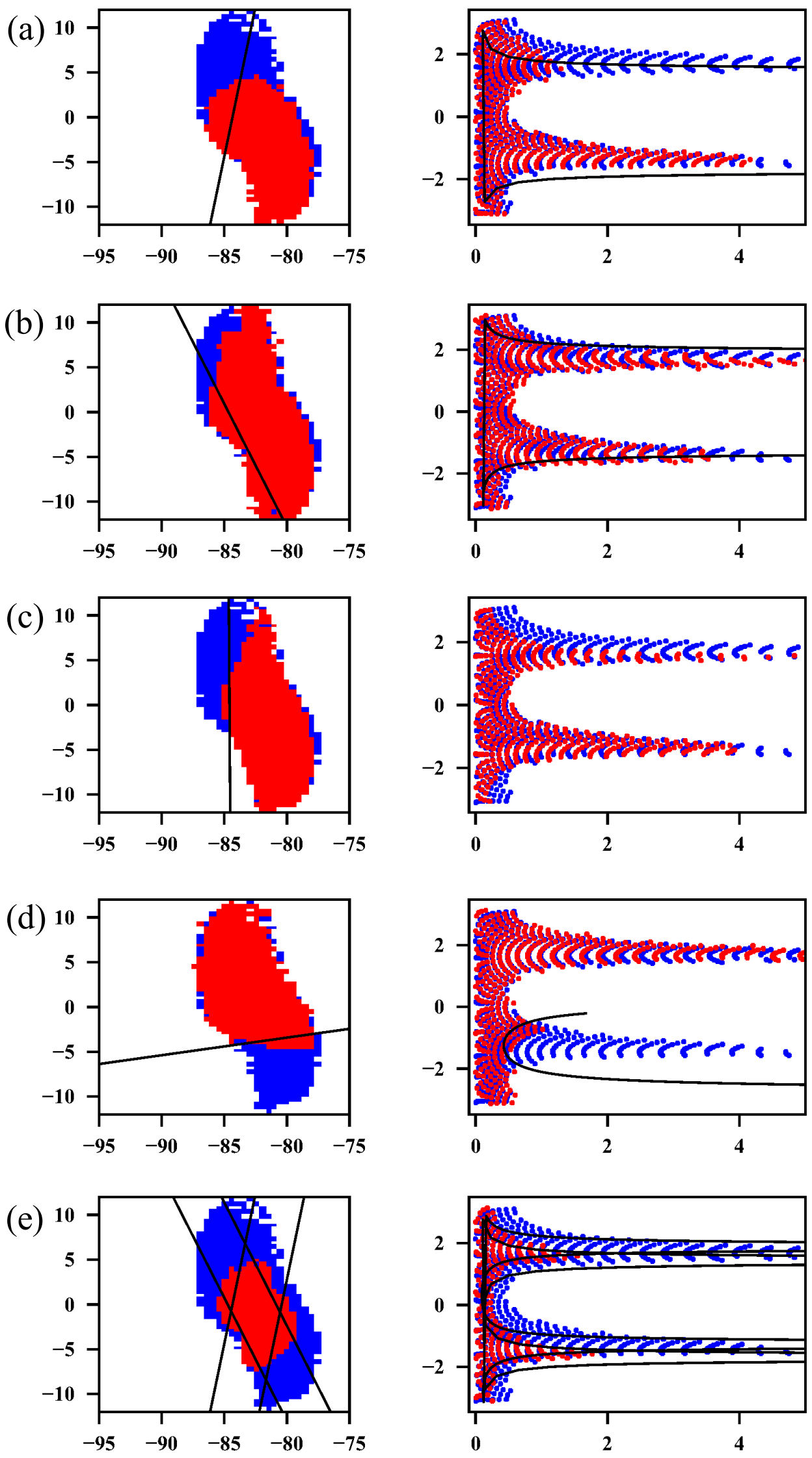

FIG. 16. Phase portraits with scraping. Rows (a) - (d) correspond to moving into the beam one of the scrapers along the beam line presented in Fig. 5; from top to bottom M00, M11, M61, M71. In each case, 0.5 mA is intercepted out the initial 5 mA. The row (e) represents the 'flat' beam when top and bottom scrapers are inserted in M00 and M11 stations.

The black lines represent the attempt of propagating the scrape lines. See other details in the text.

beam current by different scrapers results in different decrease of the tails due to a strong dependence of the tail intensity on phase. For example, inserting the scraper M71 (Fig. 16(d)) removes primarily the tail particles, while the scraper just upstream, M61 (Fig. 16(c)) misses a significant portion of the branches and does not reduce the maximum action of the beam. Instead, in order to remove 10% of the current, the scraper removes particles with lower action. Therefore, if the maximal tail truncation at a given reduction of the output beam current were the only requirement for the scraping system, beam phasing at the scrapers could be optimized by adjusting the optics and/or scraper locations such that the phases of the branches are at 0 or $\pi$ at the scrapers (the other requirements are protection from an errant beam and compatibility with the mechanical restrictions). This demonstrates the possibility to remove the large action tails with minimal beam loss by adjusting the optics such that the phases of the branches are at 0 or $\pi$ at a scraper location. Alternatively, if If such changes to the optics are not possible, scrapers that are expected to miss most of the tails can be positioned to intercept less of the total beam current.

Visually, Fig. 16 hints that propagation through the beam line smears the scraping boundary beyond of what is expected from the finite width of the scanner slits for the upstream scrapers. This could be related to nonlinear space-charge fields, as expected from simulations in Ref. [19]. In attempting to make a numerical estimation of this effect, one can propagate the scraper footprint using the transfer matrix and calculate the portion of the particles beyond the cut line in the recorded portrait. A scraper with vertical offset $d$ from the beam center produces a line in the Allison scanner portrait

$$y_1'(y) = \frac{y}{\beta_1}\left(\cot(\Delta\phi) - \alpha_1\right) - \frac{d}{\sqrt{\beta_0\beta_1}\,\sin(\Delta\phi)} \qquad (10)$$

where subscripts 0 and 1 denote the locations of the scraper and Allison scanner, correspondingly, $\Delta\phi$ is the vertical betatron phase advance between them, and $\alpha$ and $\beta$ are the Twiss parameters. The center of the coordinate system is placed at the center of the distribution. The rms Twiss parameters at the scanner are measured directly; the offset and the rms beam size and, assuming a constant emittance, $\beta_0$ can be reconstructed from the corresponding scraper scan. The phase advance, however, needs to be delivered by the optics model (simulated by TraceWin [28]). The scraper footprints drawn according to Eq. 10 are shown on all plots of Fig. 16. These lines were expected to approximately coincide with to the scraped edge of the beam distribution. Unfortunately, this visually is not the case, and numerical estimations of the particles' diffusion over the scraper footprints cannot be made. One of speculations is that the betatron phase advance experienced by the tails differs significantly from the core one, which is provided by the simulations. More accurate comparisons with simulations are expected to be performed in the future.

Another application of the scrapers is to create a low-emittance beam for initial tuning or special measurements. For example, to test the MEBT chopper [3], the four vertical scrapers in the first pair of scraper assemblies were used to remove a large part of the beam. A phase portrait of this so-called "flat" beam is shown in the last row of Fig. 16. After scraping 40% of the beam, the rms normalized emittance decreases from 0.27 $\mu$m to 0.09 $\mu$m (note that in this specific measurement the beam was not tuned optimally upstream of the RFQ thus causing the emittance of the unscraped beam to be higher than the nominal 0.2 mm mrad). With this configuration of scraping, the maximum possible action is determined by the offset of each scraper from the beam centroid and the phase advance between the two scraper assemblies. The phase dependence of the tails has little effect on the maximum action due to the heavy scraping required to prepare such a beam. The maximum action, in this case, is determined by the overlap of the scraped regions in phase space.

## VII. SUMMARY

An Allison scanner has been designed, constructed, calibrated, and successfully used in the MEBT of the PIP2IT accelerator at Fermilab to measure the phase space distributions of a 2.1 MeV $H^-$ beam. A procedure to remove the signal noise in the phase portraits independently of the beam characteristics was developed and applied in the studies.

The phase portraits, recorded at various locations, scanner orientations, beam currents, and optics settings, were analyzed in action – phase coordinates. With the Twiss parameters chosen according to properties of the beam core, the intensity of the core pixels is found to be well approximated by an exponentially decaying with action, phase-independent fit. In contrast, intensity in the beam tails drops slower with action and is strongly dependent on the phase, with two branches separated by approximately $\pi$.

The beam rms emittance is measured to be the same between two planes and at three locations. Comparison of exponential fits to the beam core shows a slight but measurable difference between the planes but no change in two locations for the same (vertical) plane. Outside of the core and well-measured tails (beyond of 99% portion of the measured beam intensity), there are indications of a halo growth along the beam line.

The phase portraits are recorded with pulse lengths of up to 0.5 ms. The beam behavior in long pulses is represented well by measurements with short (10 $\mu$s) pulses with variations on the order of ±5% over a 0.5 ms pulse.

Comparison of portraits without and with scraping shows that the scraped areas in phase space stay largely particle free when the beam propagates through the beam line. Therefore, scraping in the upstream part of the MEBT is an efficient way for creating a beam with well-defined boundaries.

## ACKNOWLEDGMENTS

Authors are grateful to the Fermilab Accelerator Division's supporting teams for making any of the measurements possible, particularly to R. Andrews (mechanical integration), D. Lambert (mechanical assembly), and A. Saewert (electrical support). The Allison scanner data acquisition program was inherited from SNS and modified by D. Slimmer. C. Richard is thankful to S. Lidia for his scientific mentoring.

This manuscript has been authored by Fermi Research Alliance, LLC under Contract No. DE-AC02-07CH11359 with the U.S. Department of Energy, Office of Science, Office of High Energy Physics.

Work is supported by the US Department of Energy, Office of Science, High Energy Physics under Cooperative Agreement award number DE-SC0018362 and Michigan State University.

### Appendix A: Calculation of the noise removal threshold for phase portrait analysis

Let us consider a rectangular portrait containing $N_{pixels} = K \times M$ pixels for which intensities are determined by random noise so that the probability density $P_p$ of finding a pixel with a given intensity $I_p$ is

$$\frac{dP_p}{dI_p} = \frac{1}{\sqrt{2\pi}\sigma_n} e^{-\frac{I_p^2}{2\sigma_n^2}}. \tag{A1}$$

The probability $P_0$ of having a pixel with intensity $A_n$ times higher than the rms noise amplitude $\sigma_n$ is

$$P_0 = 0.5\mathrm{erfc}\left(\frac{A_n}{\sqrt{2}}\right). \tag{A2}$$

The probability $P_1$ of having at least one pixel above the threshold is

$$P_1 = 1 - (1 - P_0)^{N_{pixels}} \approx P_0 N_{pixels}. \tag{A3}$$

If the cleaning procedure employed removes all the lone pixels above the threshold, spots with noise remain only if the intensities of a pair of neighboring pixels (side-by-side or diagonally) fluctuates above $A_n \sigma_n$. The total number of independent neighboring pairs $N_{pairs}$ is

$$N_{pairs} = 4KM - 3(K + M) + 2, \tag{A4}$$

which tends to

$$N_{pairs} \approx 4N_{pixels} \text{ for } K \gg 1,\ M \gg 1. \tag{A5}$$

The probability $P_2$ that two neighboring pixels are both above the threshold is

$$P_2 = 1 - (1 - P_0^2)^{N_{pairs}} \approx 4P_0^2 N_{pixels}. \tag{A6}$$

Thus, for a given $P_0$, the procedure of removing loners decreases the probability that noise pixels remain by $\sim 4P_0$. In practice, we accept that one in $\sim$ 100 portraits may contain an un-removed noise pair ($P_2 = 0.01$) and calculate the value of $P_0$ from Eq. (A4), (A6) and then the threshold by inverting Eq. (A2). For a typical number of pixel of 3000, the multiplier in Eq. (1) is $A_n = 2.5$. Without the removal of loners, it would be 3.8 for the same frequency of occurrence of portraits with un-removed noise pixels.